\documentclass[preprint,aps]{revtex4}
\usepackage{euscript}
\usepackage{amsmath}
\usepackage{epsfig}
\begin{document}
\preprint{UCI-TR-2009-16}
\title{Noninteracting Fields in the Higher Dimensional Super conducting Cosmic String Field Source Model}
\author{Aaron Roy\footnote{Electronic address:roya@uci.edu}
}
\affiliation{
Department of Physics and Astronomy, University of California, Irvine,
California 92697-4575}

\begin{abstract}
In this paper we investigate the freely propagating fields behind the results in [6]. In [6] we calculated the magnetic dipole moment of the muon and the electric dipole moments of the muon, electron and the neutron (in a simple quark model) to one loop order in both S$_1$ and S$_1 \backslash \, \boldsymbol{Z}_2$. In the analysis in [6] we took into account the effect of fields possibly generated by higher dimensional super conducting cosmic strings [4] that interact with the charged matter fields on the UED manifold. In comparing the results in [6] with standard model precision tests for the electric and magnetic dipole moments of the various fermions in the model, we were able to obtain upper limits on the compactification size as well as an upper limit for the new $b$ parameter. In [5] we presented the full theory for [6] in $M_4 \, \bigotimes$ S$_1 \backslash \, \boldsymbol{Z}_2$ where, in [6], we allowed for external magnetic fields, that could be produced by light charged particles traveling near super conducting cosmic strings, to permeate the extra dimensional space. These fields affected the charged particles in the model resulting in a novel mechanism for parity violation in QED processes as well as new mechanism for SU(2)$\bigotimes$U(1) symmetry breaking along with other phenomenological and theoretical implications [6].      
\end{abstract}

\maketitle

\section{Introduction}
For light charged particles traveling near super conducting cosmic strings [4] with a separation scale on the order of the compactification size of S$_1 \backslash \, \boldsymbol{Z}_2$, magnetic fields can be produced which will intern have a flux associated with our UED manifold $M_4 \, \bigotimes$ S$_1 \backslash \, \boldsymbol{Z}_2$. These fields will interact with the particle fields confined to the manifold. In the limit of small compactification size for our geometry, the dominant portion of the magnetic flux will come from field lines in the direction perpendicular to the extra coordinate on S$_1 \backslash \, \boldsymbol{Z}_2$ (if we think of the manifold $M_4 \, \bigotimes$ S$_1 \backslash \, \boldsymbol{Z}_2$ as being a right cylinder). In this limit, the model will be a total of six dimensions, the 5-D manifold where the particles of the model are confined, and an extra spatial dimension for the flux lines perpendicular to the extra coordinate. The flux would then be, in this limit, approximately a constant in time and proportional to the area mapped out by the extra coordinate in S$_1 \backslash \, \boldsymbol{Z}_2$. In this sense, we are only concerned with the affect of the fluxes on the particles of the model confined to $M_4 \, \bigotimes$ S$_1 \backslash \, \boldsymbol{Z}_2$ and not on the dynamics of the cosmic strings themselves which can be quite complicated in 6-D [7],[8].  

In general, the fluxes will give the charged fields of the model nontrivial periodicities, where these periodicities are not simply a shift in mode number for the field modes. There will be a new parameter introduced for the full EW spectrum in SU(2)$\bigotimes$U(1), this will be denoted as the flux parameter $b$. The introduction of these fluxes will produce a new way for EW symmetry breaking as well as a new mechanism for parity violation in QED [6]. The fluxes also induce nontrivial couplings of the Higgs and fermion fields to the gauge fields as well as different masses for the charged particles in the model for the zero modes that differ from the standard model. 

Since this model violates parity in QED, the pertinent diagrams in [6] will have contributions to the EDM's of the charged fermions in the model. These contributions were calculated in [6] along with the magnetic dipole moment contributions and we were able to find smaller upper limits for the muon EDM than the current limits in [9] as well as competitive constraints on the upper limit of the compactification size [6]. It should be noted that the contributions from the Higgs and weak gauge diagrams were neglected in [6], where we only considered the QED contributions. The reason for this is because, even with the fluxes present, a suppression factor of $(m_{\rm{lepton}}/M_W)^2$ is present for the gauge boson diagram contributions to the anomalous magnetic dipole moment that is not present in the QED diagram contributions for one extra dimension (equation (48) of [10]).             

For very large times, a fully interactive EW model in $M_4 \, \bigotimes$ S$_1 \backslash \, \boldsymbol{Z}_2$ will look like a series of field modes propagating on their own without any interactions. In our full model, it is important to know the exact expressions for the freely propagating fields and more importantly, what are their masses as a result of the external fluxes associated with our geometry. These free fields and their masses are important phenomenologically. We will first have to write out the Lagrangian density for our system up to quadratic order in the fields. Then we will integrate out the extra coordinate by forming the action for the model. With our effective 4-D Lagrangian density in terms of the field modes, we will then have to, rather carefully, redefine the gauge fields such that all interaction terms are eliminated up to quadratic order. The mass terms can then simply be determined from the quadratic terms of these freely propagating fields from our effective 4-D model. There will also be unwanted fifth components of the gauge fields, however these do not couple to any of the other fields in the model due to the orbifold geometry in as detailed in [5].

In Sec. II the general theory of the model in [5] will be summarized. In Sec. III we will write the 5-D Lagrangian density out to quadratic order and then factor the fields in a convenient way. In Sec. IV the extra coordinate will be integrated out and further factoring of the result to make the field redefinitions more transparent. Finally in Sec. V we will present the expressions for the freely propagating fields and their masses for the model followed by a brief discussion of the results. The unphysical 5th components of the gauge fields for the model were dealt with in [5] using an orbifold geometry and thus do not couple to any of the other fields in the model.      
\section{Summary of the general theory}
In the full SU(2)$\bigotimes$U(1) theory in [5] we had 
\begin{equation}
\EuScript{L} = (D_A \varphi )^\dagger (D^A \varphi) - \frac{1}{2}Tr(F_{AB}F^{AB}) - \frac{1}{4}f_{AB}f^{AB} + \mu ^2 \varphi ^\dagger \varphi - \frac{\lambda}{2}(\varphi ^\dagger \varphi )^2
\end{equation}
where $A=0,1,2,3,5$ with each field being a function of $x^\mu ,y$ where $y$ is the extra coordinate in our system and is taken as an arc length. Here the metric assumed is 
\begin{equation}
g^{AB}=
  \begin{cases} 
    0& \text{if $A\neq B$},\\
   -1& \text{if $A=B=1,2,3,5$},\\ 
    1& \text{if $A=B=0$}. 
  \end{cases} 
\end{equation}
The charged fields are affected by a factor $e^{iQby/R}$ where $b = \frac{e}{\hbar c}\times (\rm {flux})$ and $Q$ is the charge of the field ($Q = -1$ for the electron etc.). It is simplest to work in units $\hbar = c = 1$.
Then with the external flux, 
\begin{equation}
\phi \underset{\rm{flux}}{\longrightarrow}B\phi 
\end{equation}
where $B = \begin{pmatrix} e^{iby/R} & 0 \\ 0 & 1\end{pmatrix}$ since the top component of $\phi$ has $Q = +1$ and the bottom component is neutral in the standard SU(2)$\bigotimes$U(1) EW scheme. We also have
\begin{equation}
W_A = W_A ^i \frac{\tau ^i}{2} = \begin{pmatrix} \frac{1}{2}W_A ^3 & \frac{1}{\sqrt{2}}W_A ^+ \\ \frac{1}{\sqrt{2}}W_A ^- & - \frac{1}{2}W_A ^3 \end{pmatrix}\, ,  
\end{equation}
and so 
\begin{equation} 
W_A \underset{\rm{flux}}{\rightarrow} \\ \begin{pmatrix} \frac{1}{2}W_A ^3 & \frac{1}{\sqrt{2}}e^{iby/R}W_A ^+ \\ \frac{1}{\sqrt{2}}e^{-iby/R}W_A ^- & - \frac{1}{2}W_A ^3\end{pmatrix} \nonumber
\end{equation}
or more compactly,
\begin{equation}
W_A \underset{\rm{flux}}{\longrightarrow} B W_A B^\dagger \, .                                      
\end{equation}
Similarly, 
\begin{equation}
F_{AB} = \partial _A W_B - \partial _B W_A - ig[W_A , W_B].
\end{equation}
Then we have 
\begin{equation}
{\rm{Tr}}(F_{\mu \nu}F^{\mu \nu})\underset{\rm{flux}}{\longrightarrow}{\rm{Tr}}(F_{\mu \nu}F^{\mu \nu}) \, .
\end{equation}
as well as
\begin{equation}
F_{\mu 5} \underset{\rm{flux}}{\rightarrow} BF_{\mu 5}B^\dagger - (\partial _y B) W_\mu B^\dagger - B W_\mu (\partial _y B^\dagger) 
\end{equation} 
or 
\begin{equation}
{\rm{Tr}}(F_{\mu 5}F^{\mu 5})\underset{\rm{flux}}{\longrightarrow}\frac{1}{2}F^3 _{\mu 5}F^{3\mu 5} - (\partial _\mu W_5 ^- - \partial _y W_\mu ^- + \frac{ib}{R}W_\mu ^- )(\partial ^\mu W_5 ^+ - \partial _y W^{\mu +} - \frac{ib}{R}W^{\mu +}) \nonumber
\end{equation}
\begin{equation}
+ \rm{(cubic \, and \, quartic \, terms)}. 
\end{equation}
We can see that the charged $W$'s pick up a mass due to the fluxes and as a result, the SU(2)$\bigotimes$U(1) symmetry is broken. Finally with the field redefinitions in [5] required to solve the degrees of freedom problem (the $W$'s pick up a mass from the external flux before the Higg's mechanism),
\begin{equation}
\tilde{W}^- _\mu = W^- _\mu - \Lambda \partial _\mu W_5 ^- \, ,
\end{equation}
\begin{equation}
\tilde{W}^+ _\mu = W^+ _\mu - \beta \partial _\mu W_5 ^+ \, ,
\end{equation} 
we then had 
\begin{equation}
{\rm{Tr}}(F_{\mu 5}F^{\mu 5})\underset{\rm{flux}}{\longrightarrow}\frac{1}{2}F^3 _{\mu 5}F^{3\mu 5}  - ( \partial _y \tilde{W}_\mu ^- - \frac{ib}{R}\tilde{W}_\mu ^- )(\partial _y \tilde{W}^{\mu +} + \frac{ib}{R}\tilde{W}^{\mu +}) \nonumber
\end{equation}
\begin{equation}
+ \rm{(cubic \, and \, quartic \, terms)} \, . 
\end{equation}
Here $\Lambda = (\partial _y - \frac{ib}{R})^{-1}$ and $\beta = (\partial _y + \frac{ib}{R})^{-1}$. For more details of the general theory, \\ please see [5]. 
\section{Freely propagating fields}
Now we want to look at our system for very large times, in which the fields should all propagate by themselves with no interactions taking place. In this limit, we need only look at the quadratic pieces of our model and then ask ourselves, what are the freely propagating fields and what are the masses for these fields? To answer this question, we have 
\begin{equation}
\EuScript{L} = (D_A \varphi )^\dagger (D^A \varphi) - \frac{1}{2}Tr(F_{AB}F^{AB}) - \frac{1}{4}f_{AB}f^{AB} + \mu ^2 \varphi ^\dagger \varphi - \frac{\lambda}{2}(\varphi ^\dagger \varphi )^2 \nonumber
\end{equation}
\begin{equation}
= \phi ^\dagger[\overleftarrow{\partial} _\mu + ig\tilde{W}_\mu + ig\partial _\mu T + \frac{i}{2}g'B_\mu][\overrightarrow{\partial} ^\mu - ig\tilde{W}^\mu - ig\partial ^\mu T - \frac{i}{2}g'B^\mu]\phi \nonumber
\end{equation}
\begin{equation} 
- \phi ^\dagger[\overleftarrow{\partial _y} + (\partial _y B^\dagger )B + igW_5 + \frac{i}{2}g'B_5][\overrightarrow{\partial _y} + B^\dagger (\partial _y B) - igW_5 - \frac{i}{2}g'B_5]\phi \nonumber
\end{equation}
\begin{equation} 
+ ( \partial _y \tilde{W}_\mu ^- - \frac{ib}{R}\tilde{W}_\mu ^- )(\partial _y \tilde{W}^{\mu +} + \frac{ib}{R}\tilde{W}^{\mu +}) - \frac{1}{2}F^3 _{\mu 5}F^{3\mu 5} - \frac{1}{2}Tr(F_{\mu \nu}F^{\mu \nu}) 
- \frac{1}{4}f_{AB}f^{AB} + \mu ^2 \varphi ^\dagger \varphi \nonumber
\end{equation}
\begin{equation}
- \frac{\lambda}{2}(\varphi ^\dagger \varphi )^2 + (\rm {cubic \, and \, quartic \, terms \,}). \nonumber
\end{equation}
In terms of the vacuum expectation value $v$ we then let 
\begin{equation}
\phi = \frac{1}{\sqrt{2}}\begin{pmatrix} \phi _1 \\ v+\phi _2 \end{pmatrix} \, , 
\end{equation}
which then gives, after a considerable amount algebra,  
\begin{equation}
\EuScript{L}=\frac{g^2 v^2}{4}\Bigg [ \tilde{W}^{+\mu} + \frac{i\sqrt{2}}{gv}\partial ^\mu \phi _1 
+ \beta (\partial ^\mu W_5 ^+)\Bigg ] \Bigg [ \tilde{W}^- _\mu - \frac{i\sqrt{2}}{gv}\partial _\mu \phi _1 ^* 
+ \Lambda (\partial _\mu W_5 ^-) \Bigg ] \nonumber
\end{equation}
\begin{equation}
+ \frac{v^2}{8}(g^2 + g^{'2})\Bigg [ Z^\mu - \frac{2i}{v\sqrt{g^2 + g^{'2}}}(\partial ^\mu \phi _2) \Bigg ] \Bigg [ Z_\mu + \frac{2i}{v\sqrt{g^2 + g^{'2}}}(\partial _\mu \phi _2 ^*) \Bigg ] \nonumber
\end{equation}
\begin{equation}
- \frac{1}{2}\Bigg [ \partial _y \phi _1 + \frac{ib}{R}\phi _1 - \frac{igv}{\sqrt{2}}W_5 ^+ \Bigg ] \Bigg [ \partial _y \phi _1 ^* - \frac{ib}{R}\phi _1 ^* + \frac{igv}{\sqrt{2}}W_5 ^- \Bigg ] \nonumber
\end{equation}
\begin{equation}
- \frac{v^2}{8}(g^2 + g^{'2})\Bigg [ Z_5 - \frac{2i}{v\sqrt{g^2 + g^{'2}}}(\partial _y \phi _2) \Bigg ] \Bigg [ Z_5 + \frac{2i}{v\sqrt{g^2 + g^{'2}}}(\partial _y \phi _2 ^*) \Bigg ] \nonumber
\end{equation}
\begin{equation}
+ ( \partial _y \tilde{W}_\mu ^- - \frac{ib}{R}\tilde{W}_\mu ^- )(\partial _y \tilde{W}^{\mu +} + \frac{ib}{R}\tilde{W}^{\mu +}) - \frac{1}{2}F^3 _{\mu 5}F^{3\mu 5} - \frac{1}{2}Tr(F_{\mu \nu}F^{\mu \nu}) - \frac{1}{4}f_{AB}f^{AB} \nonumber
\end{equation}
\begin{equation}
+ \mu ^2 \varphi ^\dagger \varphi - \frac{\lambda}{2}(\varphi ^\dagger \varphi )^2 + (\rm {cubic \, and \, quartic \, terms \,}). \nonumber 
\end{equation}
\section{The field modes}
Now let us integrate out the extra coordinate, keeping in mind that for the gauge fields the modes are the same fields as the anti modes. This is because equation (1) is invariant under parity even for nonzero flux, then it must also be parity invariant when we integrate out the extra coordinate and this condition forces the gauge field modes to be the same fields as the anti modes. If we add the QED portions, then the model becomes parity violating for nonzero flux. Please note that \, $\beta \rightarrow \beta _n = -\frac{iR}{n+b}$ \, and \, $\Lambda \rightarrow \Lambda _n = \frac{iR}{n+b}$ \, once we integrate out the modes. Then we have, after a considerable amount of factoring,
\begin{equation}
\EuScript{L}_{\rm{effective}}=\sum _{n\,=\,-\infty} ^\infty\Bigg \{ \frac{1}{2}\Bigg (\frac{g^2 v^2}{2} + \frac{(n+b)^2}{R^2}\Bigg )\Bigg [ \tilde{W}_n ^{+\mu} + \frac{\frac{igv}{\sqrt{2}}(\partial ^\mu \phi _{1, n}) + \frac{g^2 v^2}{2}\beta _n (\partial ^\mu W_{5, n} ^+)}{
\frac{g^2 v^2}{2} + \frac{(n+b)^2}{R^2}}\Bigg ] \times \nonumber
\end{equation}
\begin{equation}
\Bigg [ \tilde{W}_{\mu, n} ^- + \frac{\frac{-igv}{\sqrt{2}}(\partial _\mu \phi _{1, n} ^*) + \frac{g^2 v^2}{2}\Lambda _n (\partial _\mu W_{5, n} ^-)}{\frac{g^2 v^2}{2} + \frac{(n+b)^2}{R^2}}\Bigg ] \nonumber
\end{equation}
\begin{equation}
- \frac{1}{2}\frac{(\partial ^\mu \phi _{1, n} - \frac{igv}{\sqrt{2}}\beta _n \partial ^\mu W^+ _{5, n})(\partial _\mu \phi ^* _{1, n} + \frac{igv}{\sqrt{2}}\Lambda _{n} \partial _\mu W^- _{5, n})}{\frac{2}{g^2 v^2}(\frac{g^2 v^2}{2} + \frac{(n+b)^2}{R^2})} \nonumber 
\end{equation}
\begin{equation}
+ \frac{1}{2}(\partial ^\mu \phi _{1, n} - \frac{igv}{\sqrt{2}}\beta _n \partial ^\mu W^+ _{5, n})(\partial _\mu \phi ^* _{1, n} + \frac{igv}{\sqrt{2}}\Lambda _{n} \partial _\mu W^- _{5, n}) \nonumber
\end{equation}
\begin{equation} 
+ \frac{v^2}{8}(g^2 + g^{'2})\Bigg [ Z^\mu _n - \frac{2i}{v\sqrt{g^2 + g^{'2}}}(\partial ^\mu \phi _{2, n}) \Bigg ] \Bigg [ Z_{\mu, n} + \frac{2i}{v\sqrt{g^2 + g^{'2}}}(\partial _\mu \phi _{2, n} ^*) \Bigg ] \nonumber
\end{equation}
\begin{equation} 
- \frac{1}{2}\frac{(n+b)^2}{R^2}( \phi _{1, n} - \frac{igv}{\sqrt{2}}\beta _n W_{5, n} ^+)( 
\phi _{1, n} ^* + \frac{igv}{\sqrt{2}}\Lambda _n W_{5, n} ^-) \nonumber 
\end{equation}
\begin{equation}
- \frac{v^2}{8}(g^2 + g^{'2})\Bigg [ Z_{5, n} + \frac{2n}{vR\sqrt{g^2 + g^{'2}}} \phi _{2, n} \Bigg ] \Bigg [ Z_{5, n} + \frac{2n}{vR\sqrt{g^2 + g^{'2}}} \phi _{2, n} ^* \Bigg ] - \frac{1}{2}F^3 _{\mu 5, n}F^{3\mu 5}_{-n} \nonumber 
\end{equation}
\begin{equation}
- \frac{1}{2}Tr(F_{\mu \nu , n}F^{\mu \nu} _{-n}) - \frac{1}{4}f_{AB, n}f^{AB} _{-n} + \mu ^2 \varphi _n ^\dagger \varphi _n - \frac{\lambda}{2}(\varphi _n ^\dagger \varphi _n )^2 \Bigg \} + (\rm {cubic \, and \, quartic \, terms \,}). \nonumber
\end{equation}
Please note that 
\begin{equation}
F^3 _{\mu 5,n}F^{3\mu 5} _{-n} = -(\partial _\mu W^3 _{5,n} - \frac{in}{R}W^3 _{\mu ,n})(\partial ^\mu W^3 _{5,-n} + 
\frac{in}{R}W^{3,\mu} _{-n}) \nonumber
\end{equation}
and
\begin{equation}
f _{\mu 5,n}f^{\mu 5} _{-n} = -(\partial _\mu B _{5,n} - \frac{in}{R}B _{\mu ,-n})(\partial ^\mu B _{5,-n} + 
\frac{in}{R}B^\mu _{-n}) \, . \nonumber
\end{equation}
\section{Field redefinitions}
Let us define the following fields: 
\begin{equation}  
\tilde{Z} _{5,n} = Z_{5,n} + \frac{2n}{vR\sqrt{g^2+g^{'2}}}\rm{Re}\phi _{2,n} \, ,
\end{equation}
\begin{equation}
\tilde{Z} _{\mu ,n} = Z_{\mu ,n} + \frac{2}{v\sqrt{g^2+g^{'2}}}\rm{Im} (\partial _\mu \phi _{2,n}) \, ,
\end{equation}
\begin{equation}
W^{'+\mu} _n = \tilde{W}_n ^{+\mu} + \frac{\frac{igv}{\sqrt{2}}(\partial ^\mu \phi _{1, n}) + \frac{g^2 v^2}{2}\beta _n (\partial ^\mu W_{5, n} ^+)}{\frac{g^2 v^2}{2} + \frac{(n+b)^2}{R^2}} \, ,
\end{equation}     
\begin{equation} 
\chi _n = \phi _{1, n} - \frac{igv}{\sqrt{2}}\beta _n W^+ _{5, n} \, .
\end{equation} 
With these new field definitions we can diagonalize the effective 4-D Lagrangian density up to quadratic order. Substituting these field redefinitions, which are the freely propagating fields for the model, we then have finally 
\begin{equation}
\EuScript{L}_{\rm{effective}} = \sum _{n\,=\,-\infty} ^\infty\Bigg \{\frac{1}{2}(\frac{g^2 v^2}{2} + \frac{(n+b)^2}{R^2})
W^{'+} _{\mu , n}W^{'-\mu} _{-n} \nonumber
\end{equation}
\begin{equation}
+ \frac{1}{2}(\partial _\mu \chi _n)(\partial ^\mu \chi ^* _n)(1 - \frac{g^2 v^2}{2(\frac{g^2 v^2}{2} + \frac{(n+b)^2}{R^2})}) -\frac{1}{2}\frac{(n+b)^2}{R^2}(\chi _n)(\chi ^* _n) + \frac{1}{2}(\partial _\mu h_{-n})(\partial ^\mu h_n) \nonumber
\end{equation}
\begin{equation}
- \frac{1}{2}(4\mu ^2) h_{-n} h_n + \frac{v^2}{8}(g^2 + g^{'2})\tilde{Z} _{\mu, -n} \tilde{Z} ^\mu _n + \frac{v^2}{8}(g^2 + g^{'2})\tilde{Z} _{5, -n} \tilde{Z} ^5 _n \nonumber
\end{equation}
\begin{equation}
- \frac{1}{2}\frac{n^2}{R^2} l_{-n} l_n - \frac{1}{2}F^3 _{\mu 5, n}F^{3\mu 5}_{-n} - \frac{1}{2}Tr(F_{\mu \nu , n}F^{\mu \nu} _{-n}) - \frac{1}{2}f_{\mu 5, n}f^{\mu 5} _{-n} \nonumber
\end{equation}
\begin{equation}
-\frac{1}{4}f_{\mu \nu, n}f^{\mu \nu} _{-n} \Bigg \} + (\rm {cubic \, and \, quartic \, terms \,}). 
\end{equation}
Here $h = \rm{Re} (\phi _2)$ and $l = \rm{Im} (\phi _2)$. Notice that $\frac{1}{2}F^3 _{\mu 5, n}F^{3\mu 5}_{-n}, \, \frac{1}{2}Tr(F_{\mu \nu , n}F^{\mu \nu} _{-n})$ and $\frac{1}{4}f_{\mu 5, n}f^{\mu 5} _{-n}, \, \\ \frac{1}{4}f_{\mu \nu, n}f^{\mu \nu} _{-n}$ are invariant under the above transformations up to quadratic order when cast in the $Z$ and $\tilde{W}$ basis. 

Reading off the mass terms for the modes from equation (18) we have
\begin{equation}
m_{W'_n} = \sqrt{\frac{g^2 v^2}{2} + \frac{(n+b)^2}{R^2}}
\end{equation} 
\begin{equation}
m_{Z_n} = \sqrt{\frac{v^2 (g^2 + g'^2)}{4} + \frac{n^2}{R^2}}
\end{equation}
\begin{equation}
m_{A_n} = \frac{|n|}{R}
\end{equation}
\begin{equation}
m_{Z_{5, n}} = \frac{v}{2}\sqrt{g^2 + g^{'2}}
\end{equation}
\begin{equation}
m_{h_n} = \sqrt{4\mu ^2 + \frac{n^2}{R^2}}
\end{equation}
\begin{equation}
m_{\chi _n} = \frac{|n+b|}{R} \, .
\end{equation}
The $Z_{5, n}$ does not couple to the other modes due to the orbifold geometry and is thus phenomenologically absent from the diagrams involving the other fields in the model [5]. The mode dependence for $m_{Z_n}$ and $m_{A_n}$ comes from $- \frac{1}{2}F^3 _{\mu 5, n}F^{3\mu 5}_{-n} - \frac{1}{2}f_{\mu 5, n}f^{\mu 5} _{-n}$. Notice that there is a mass term for $l_n$, but this is not really a mass term because this field does not have a kinetic term associated with it and thus it carries no degrees of freedom of motion. It does however act as a constraint to the system such that it imparts an additional contribution to the mass of $m_{h_n}$. 

To see this, simply write $h_n$ and $l_n$ as a linear combination of two new fields such that there are no cross terms for the field derivatives. These new fields will couple to one another and the equation of motion for one of the fields will impart an additional mass contribution to the other field. This is where the mode dependence for $m_{h_n}$comes from (note that we have not written the final result in terms of the new fields for convenience, it does not matter what we call these fields anyway). The remaining unwanted mass term vanishes in the zero mode limit. It is also clear that in the zero mode and zero flux limit, equation (18) and the mass terms reduce to the standard model results once the fifth component for each gauge field is orbifolded away [5]. It should be noted however that the field redefinition of $\tilde{W}^\mu _n$ is undefined in the zero mode limit when $b=0$ just as in the general case [5].

\end{document}